\documentclass[12pt]{article}

\usepackage{epsfig}
\usepackage[inkscapelatex=false]{svg}
\usepackage{physics}
\usepackage{geometry}
\usepackage{graphicx}
\usepackage{subfig}
\usepackage{multirow}
\usepackage{multicol}
\usepackage{caption}
\usepackage{floatrow}
\usepackage{float}
\usepackage{amsfonts}
\usepackage{amsmath}
\usepackage[%
    colorlinks=true,
    pdfborder={0 0 0},
    linkcolor=red]{hyperref}
\usepackage[english]{babel}
\usepackage{cite}
\usepackage{subcaption} 
\usepackage[export]{adjustbox} 
\usepackage{cleveref} 
\usepackage{soul}

\begin{document}
\title{Pressure Dependence of Ultrafast Carrier Dynamics in Excitonic Insulator Ta$_2$NiSe$_5$}
\author{Vikas Arora$^{1,2}$, Victor S Muthu$^{1}$, Arjit Sinha$^{3}$, Luminita\  Harnagea$^{4}$,\\ U V Waghmare$^{3}$, A\ K\ Sood$^{*1,2}$}
\date{\today}
\maketitle

\begin{center}
\textit{$^{1}$Department of Physics, Indian Institute of Science, Bangalore 560012, India\\
$^{2}$Center for Ultrafast Laser Applications, Indian Institute of Science, Bangalore 560012, India\\
$^{3}$Theoretical Sciences Unit, Jawaharlal Nehru Centre for Advanced Scientific Research, Bangalore 560064, India \\
$^{4}$Department of Physics, Indian Institute of Science Education and Research, Pune, Maharashtra 411008, India}
\end{center}

\begin{center}
\end{center}

\begin{abstract}
An excitonic insulator (EI) phase is a consequence of collective many-body effects where an optical band gap is formed by the condensation of electron-hole pairs or excitons. We report pressure-dependent optical pump optical probe spectroscopy of EI Ta$_2$NiSe$_5$ in an on-site in situ geometry. The fast relaxation process depicts the transition across P$_{C_1}\sim$ 1 GPa from EI phase to a semiconductor and P$_{C_2}\sim$ 3 GPa from a semiconductor to a semimetallic phase. The instability of the EI phase beyond P$_{C_1}$ is captured by the Rothwarf-Taylor model by incorporating the decrease of the bandgap under pressure. The pressure coefficient of the bandgap decreases, 65 meV/GPa closely agrees with the first principle calculations.

\end{abstract}

\section*{Introduction}
The minimum energy required to form an exciton in a semiconductor is the difference of bandgap, E$_G$, and the binding energy of an exciton, E$_B$\cite{Knox1963,Halperin1968}. When E$_B$ exceeds E$_G$, the ground state of the material displays spontaneous formation of excitons. For semimetals exhibiting low carrier density, the presence of weakly screened Coulomb interaction results in the formation of a bound state between a hole and an electron, known as an exciton.\cite{Mott1961,Halperin1968}. The condensation of non-conducting excitons in the ground state gives rise to an insulating phase in both types of materials, which is consequently termed excitonic insulators (EI)\cite{Mott1961,Jerome1967,Halperin1968,Bucher1991}. In simpler terms, the process of excitonic condensation results in the creation of an optical gap. As depicted in a qualitative illustration of the phase boundary of an EI (Figure 1 of Ref. \cite{Bronold2006,Lu2017}), the transition from a semimetal to an EI is akin to a Bardeen–Cooper–Schrieffer (BCS)-like transition, while the transition from a semiconductor to an EI is marked by a Bose-Einstein condensate (BEC) transition. The EI phase of the material, predicted about six decades ago\cite{Mott1961,Jerome1967,Halperin1968}, has been identified in TmSe$_{0.45}$Te$_{0.55}$\cite{Bucher1991,Wachter2004}, 1\textit{T}-TiSe$_2$\cite{Cercellier2007}, Ta$_2$NiSe$_5$\cite{Wakisaka2009}, and more recently in monolayer WTe$_2$\cite{Jia2022} and twisted heterostructure of monolayer WSe$_2$ and bilayer WSe$_2$\cite{Chen2022}. In this context, the EI phase in Ta$_2$NiSe$_5$(TNSe) is remarkable for its distinct presence at room temperature.
\par
Di Salvo \textit{et al} has reported a semiconductor to EI phase transition at T=328 K\cite{DiSalvo1986} in TNSe along with orthorhombic to monoclinic structural change on cooling\cite{Kaneko2012}. TNSe exhibits a layered structure, wherein tantalum and nickel are organized in one-dimensional chains within each layer\cite{Sunshine1985}. Electron-hole Coulomb interactions hybridize Ni 3d-Se 4p valence band and Ta 5d conduction band, which characterize the ground state of TNSe with Coulomb coupling between holes and electrons at $\Gamma$ point\cite{Wakisaka2009,Wakisaka2012}. Wakisaka \textit{et al} have shown the valence band flatness of TNSe below T=328 K using angle resolved photoemission spectroscopy\cite{Wakisaka2009,Wakisaka2012}. The resistivity measurements show a kink at T=326 K for TNSe\cite{DiSalvo1986,Lu2017}. Raman active phonon modes of TNSe show splitting of 119 cm$^{-1}$ phonon mode accompanied by emergence of two new phonon modes, namely 147 cm$^{-1}$ and 235 cm$^{-1}$ in the EI state \cite{Yan2019}. Temperature-dependent X-ray diffraction (XRD) shows TaSe$_6$ octahedra nearly static, while NiSe$_4$ tetrahedra vibrations amplify at lower temperatures below T=328 K, leading to strong electron-phonon coupling in TNSe. This, coupled with electron-hole interactions, induces a bandgap in the EI state\cite{Yan2019}. Below the transition temperature, the reduced distance between Ni and Ta atoms leads to excitons with more binding energy\cite{Wakisaka2012}. Recent experiments and theoretical calculations propose that the EI phase transition in TNSe is primarily structural, driven by spontaneous symmetry breaking due to phonon instabilities\cite{Baldini2023,Windgätter2021}. The stabilization of EI phase in ultrathin (5 layered) TNSe is affected more by Coulomb interaction in comparison to electron-phonon coupling\cite{Kim_2016}. The microscopic understanding of the process involved in the transition is disclosed by ultrafast spectroscopy, where Werdehausen \textit{et al} claimed the TNSe to be located in BEC-BCS crossover regime\cite{Werdehausen2018}. In another work, they presented the evidence of coherent amplitude mode in TNSe, analogous to Higgs mode in superconductors\cite{Werdehausen2018r}. Whether strong photoexcitation with NIR pulses can melt the collective order of the EI phase is not settled, Hope et al. \cite{Bretscher2021} in favor, whereas Mor et al. \cite{Mor2018} against. Exploring a myriad of temperature-dependent phenomena, including intriguing electron-phonon coupling and coherent phonon modes in TNSe, pressure serves as another valuable tool for investigation. 
\par
Exciton formation, influenced by carrier density, can be studied by manipulating electronic density states, such as through physical pressure to adjust the band-gap width\cite{Nakano2018}. Moreover, TNSe possesses a layered structure featuring weak van der Waals interactions between individual layers compared to strong intralayer covalent bonding, making external pressure exploration particularly significant. High-pressure XRD studies of TNSe reveal structural phase transition across P=3 GPa due to the coherent sliding of layers, which originate from a shift in interlayer Se ion arrangement. It is reflected in an abnormal reduction in unit cell volume and Se-Se atom distance across the transition pressure\cite{Nakano2018}. Resistivity measurements under pressure at different temperatures unveil the phase diagram of TNSe\cite{Lu2017,Nakano2018}, which elucidates its superconducting phase at a pressure P=7.9 GPa and critical temperature T$_{SC}$ of 1.2 K\cite{Matsubayashi2021}. The phase diagram depicts that, at room temperature, TNSe undergoes a transition from an excitonic insulator (EI) to a semiconductor at P=1 GPa, and from a semiconductor to a semimetal at P=3 GPa\cite{Nakano2018}. This transition is further supported by high-pressure Raman spectroscopy findings by Sukanya \textit{et al.}\cite{SukanyaPal2020}. To comprehend the microscopic origin of these transitions, we present optical pump optical probe spectroscopy of TNSe at high pressures under room temperature conditions. We corroborate the pressure-induced transitions by changes in the relaxation times and amplitudes as a function of pressure. A linear drop in the bandgap of TNSe from ambient pressure to P$_{C_1}$ calculated theoretically captures the trend of the amplitude and the relaxation time within the Rothwarf-Taylor model\cite{Rothwarf1967,Kabanov1999,Kabanov2005}. A significant drop in the relaxation time beyond P=3 GPa illustrates the semimetallic behavior of TNSe. This study not only facilitates the ultrafast control of the excitonic insulating state at room temperature but also enhances our comprehension of the many-body effects analogous to superconductors under room temperature conditions.


\section*{Experimental details}
Ta$_2$NiSe$_5$ (TNSe) single crystals were grown by the chemical vapor transport technique, employing iodine as the transporting agent\cite{SukanyaPal2020}. A Ti: Sapphire pulsed laser (M/s Newport Corporation Pvt Ltd) with a pulse width of $\sim$50 fs and repetition rate of 1 kHz was used to perform optical pump-optical probe spectroscopy. The pump and probe lasers with photon energy of 1.55 eV were maintained in cross-polarized configuration. Pressure dependent studies were performed at two different fluences of 0.8 and 1.1 mJ/cm$^2$. The relaxation dynamics under high pressure were investigated using a symmetric diamond anvil cell (DAC). A stainless steel gasket with a pre-existing indentation and a hole of 200 $\mu$m in diameter was positioned between the diamond anvils, serving as the sample chamber. A small piece of crystalline Ta$_2$NiSe$_5$ sample (size $\approx$100 $\mu$m) was embedded in the sample chamber along with a ruby chip for pressure calibration with methanol-ethanol (4:1) mixture as a pressure-transmitting medium. 
Ultrafast measurements were conducted from the front side of the DAC, while the ruby fluorescence was recorded from the opposite side of the DAC. The ruby fluorescence measurements were carried out using a continuous-wave He-Ne laser with a wavelength of 632 nm, along with a spectrometer attached with a liquid nitrogen-cooled CCD (iHR320 by M/s Horiba Jobin Yvon).
\par
\begin{figure}[H]
   \centering
   \includegraphics[width=7.2in]{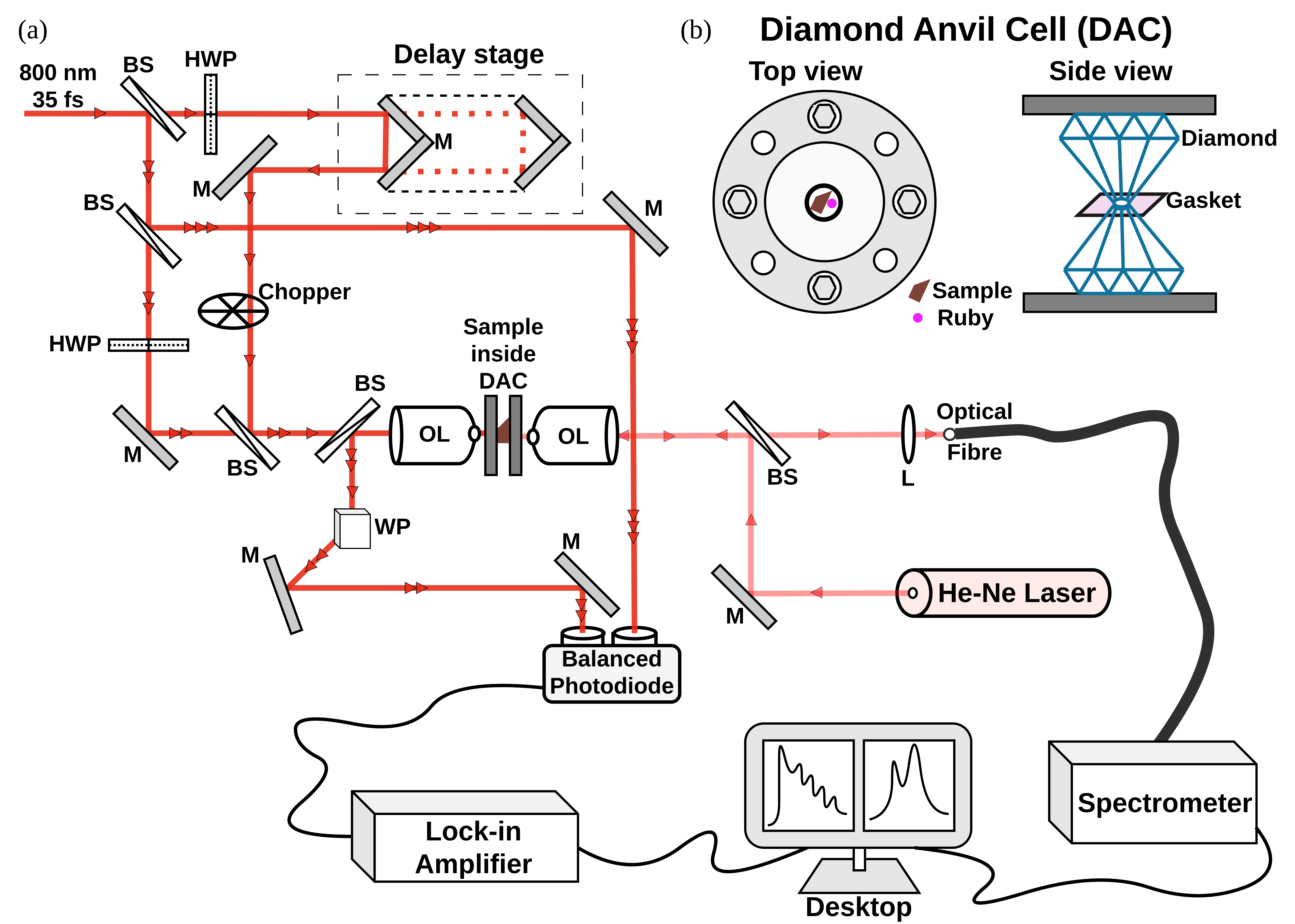}
    \caption[Experimental set-up for in-situ ultrafast spectroscopy of Ta$_2$NiSe$_5$ under pressure]{(a) With an On-site in situ geometry, a high-pressure experimental setup for ultrafast optical pump optical probe spectroscopy. Ultrafast spectroscopy measurements are depicted in the left half section (with dark red beam), whereas the pressure calibration setup for recording the ruby fluorescence is demonstrated in the right section (bottom half, with faint red beam), (b) The zoomed-out version of `Sample inside DAC' where Diamond Anvil Cell (DAC) with sample loaded inside (top and side view) is depicted. HWP: Half Wave Plate, BS: Beam Splitter, M: Mirror, OL: Objective Lens, WP: Wollaston Prism, L: Lens.}
    \label{Fig:4p1}
\end{figure}

The experimental framework for the pressure-dependent optical pump optical probe (OPOP) spectroscopy (in an on-site in situ geometry) is illustrated in Figure \ref{Fig:4p1}\textcolor{red}{a}. The OPOP measurements are represented by the bold red beam path in the left half section of the figure. The pump beam (single arrow) and the probe beam (double arrow) are aligned to be collinear, passing through an objective lens (OL) before striking the sample (inside DAC). The reflected beams retrace the path through the objective lens (OL) until reaching the first beam splitter (BS), where they are subsequently guided through a Wollaston prism (WP). The cross-polarized pump and probe beams diverge spatially with the use of a Wollaston prism (WP), and eventually, the latter is directed towards a balanced photodetector. Simultaneously, a reference beam (triple arrow) is collected at the same photodetector. The photoresponse acquired is amplified through a lock-in amplifier, and the data is then recorded in the system using the LabVIEW interface. A CCD camera, placed before the WP and utilizing an additional beam splitter (not shown here), is deployed to image the sample chamber. This camera is positioned perpendicularly to the plane of beam propagation.
\par
A top and side view of a symmetric diamond anvil cell (DAC) is displayed in the upper-right portion of Figure \ref{Fig:4p1}\textcolor{red}{b}. To adjust the pressure, the two screws of the symmetric DAC are turned clockwise, while the other two are turned counterclockwise, arranged alternately. This procedure is accomplished without removing the DAC from the beam path, elucidating the on-site, in situ pressure-dependent optical pump optical probe spectroscopy.

\section*{Computation Details}

Our first-principles calculations for estimating the bandgap as a function of pressure are based on density functional theory (DFT) as implemented in Quantum ESPRESSO package \cite{Giannozzi_2009,Giannozzi2009,Giannozzi_2017}. Interactions between valence electrons and ionic cores are modeled using projector-augmented-wave (PAW) pseudopotentials generated by Dal Corso \cite{Corso_2015}. We used Perdew-Bruke-Ernzerhof (PBE) functional of exchange-correlation energy of electrons with a generalized-gradient approximation (GGA) \cite{Perdew_1996}. Energy cut-offs of 45 Ry and 320 Ry were used to truncate plane wave basis-sets used to represent Kohn-Sham (KS) wave functions and charge density respectively. Brillouin zone (BZ) integrations were sampled on a uniform mesh of $8 \times 8 \times 3$ \textbf{k}-points in calculations of $C2/c$ structure with a Fermi-Dirac smearing width of $k_{\rm B}T$ = 0.003 Ry. For accurate description of the van der Waals (vdW) interaction in the material \cite{Subedi_2020} a non-local (nl) optB88-vdW \cite{Klimes_2009,Thonhauser_2015,Thonhauser_2007,Berland_2015,Langreth_2009,Sabatini_2012} functional is used to obtain a correction in the correlation energy. Self-consistent numerical solution of the Kohn-Sham equation was obtained with the convergence of total energy within $10^{-8}$ Ry. Structures were relaxed by minimizing total energy using the Broyden-Fletcher-Goldfarb-Shanno (BFGS) algorithm, until (a) Hellman-Feynman forces on each atom converge within $10^{-4}$ Ry/Bohr, and (b) each component of the stress tensor of the unit cell becomes less than 0.2 GPa. Since DFT typically overestimates electronic bandwidth of $d$-orbital bands, and Ta, and Ni have partially filled $d$-orbitals. We used on-site Hubbard-$U$ parameter \cite{Anisimov_2015, Cococcioni_2005}, $U$ = 6.0 eV only at Ta-$5d$ orbitals \cite{SukanyaPal2020} to capture the correct electronic band gap, because we find band gap and lattice parameters are insensitive to $U$ value at Ni-$3d$ orbitals \cite{Sukanya_2024}.

\section*{Results and discussions}
\begin{figure}[H]
  \centering
  \includegraphics[width=5in]{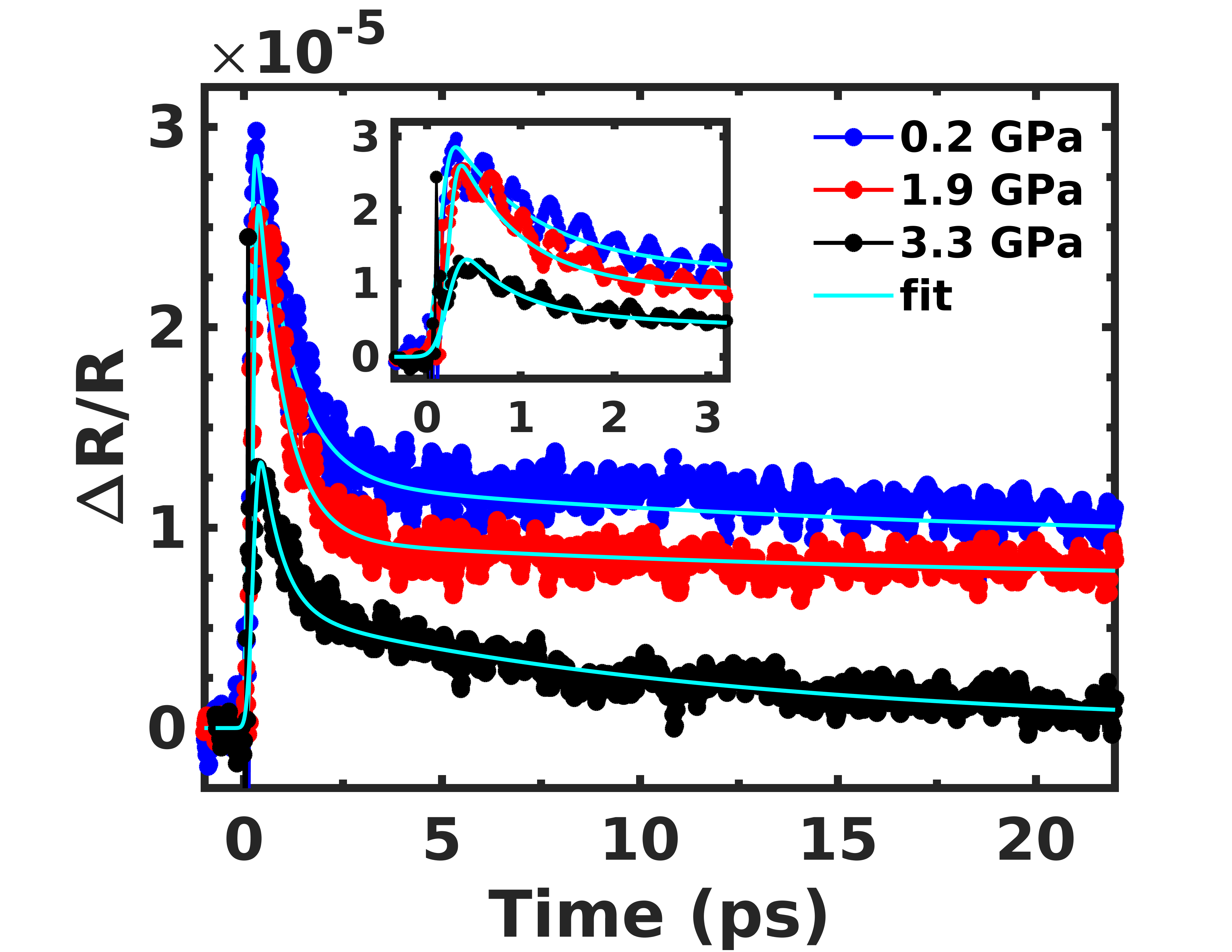}
  \caption[Time-resolved differential reflectivity  at three different pressures]{(a) Time-resolved differential reflectivity at three different pressures, P=0.2, 1.9, 3.3 GPa. Inset shows the coherent oscillations in differential reflectivity for the first 3.1 ps delay time. The solid cyan curves represent the biexponential fit (Eq. \ref{eq:4p1}) to the data (solid circles).} 
\label{Fig:4p2}
\end{figure}

Figure \ref{Fig:4p2} shows the time-resolved differential reflectivity ($\Delta R/R$) at three different pressures of 0.2, 1.9, and 3.3 GPa depicting pressure dependence of the relaxation dynamics. The inset represents the differential reflectivity ($\Delta$R/R) for the first 3.1 ps, where coherent phonon oscillations are also manifested. Electronic relaxation involves fast and slow decay processes, along with a constant long-lasting component, thus a biexponential decay model is suitable for fitting real-time data.
\begin{equation}
    \centering
    \frac{\Delta R}{R}(t)=\frac{1}{2}\left(1+erf\left(\frac{t-t_0}{\tau_{rise}}\right)\right)\left(A_1e^{-t/\tau_1}+A_2e^{-t/\tau_2}+A_3\right)
    \label{eq:4p1}
\end{equation}
where A$_1$ (A$_2$) and $\tau_1$ ($\tau_2$) represent the amplitude and relaxation time for the fast (slow) dynamical process respectively. A$_3$ represents the long-lived Bolometric signal, which illustrates the propagation of heat out of the pumped area, exhibiting a constant value on 100 ps time scale\cite{Demsar2006,Werdehausen2018}. The validity of the model is demonstrated in Figure \ref{Fig:4p2}, where the experimental data for $\Delta$R/R (represented by solid circles) is well-fitted with the biexponential decay (shown as solid cyan curve). The OPOP experiments for TNSe were done up to a pressure of 4.7 GPa, beyond which, the signal to noise ratio is too poor to extract quantitative behavior.
\par
\begin{figure}[H]
    \centering
    \includegraphics[width=\textwidth]{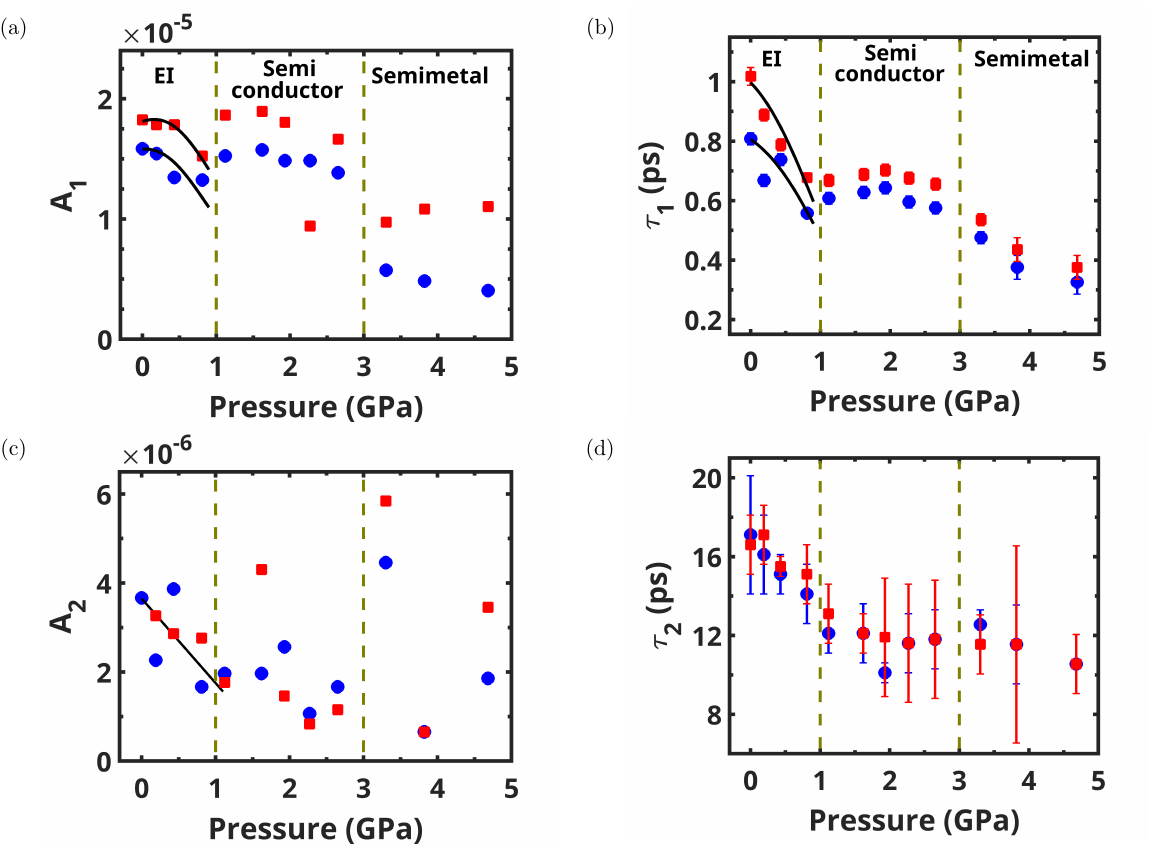}
    \caption[Relaxation dynamics parameters as a function of pressure for EI Ta$_2$NiSe$_5$]{Fitting parameters of the biexponential fit to the time-resolved differential reflectivity at different pressures for two fluences of 0.8 (blue solid circles) and 1.1 mJ/cm$^2$ (red solid circles), (a) Amplitude A$_1$ and (b) relaxation time $\tau_1$. The solid black lines in (a) $\&$ (b) are fit to the RT model. (c) Amplitude A$_2$, (d) Decay time $\tau_2$. The black solid lines in (c) and (d) are guides to the eye. The error bars for the amplitudes are smaller than the size of the blue (red) circle (square).}
    \label{Fig:4p3}
\end{figure}

Figure \ref{Fig:4p3} displays the fitting parameters (A$_1$,$\tau_1$) and (A$_2$,$\tau_2$) at different pressures for two distinct fluences of 0.8 (blue circles) and 1.1 mJ/cm$^2$ (red squares). The amplitude (A$_1$) of the fast relaxation process exhibits a gradual decline in the EI region (from P=0 to 1 GPa), shows an increase at the transition pressure and slightly decreases in the semiconductor region (from P=1 to 3 GPa), and undergoes an abrupt decrease in the semimetal region (beyond P=3 GPa)\cite{Nakano2018,SukanyaPal2020}. The corresponding relaxation time ($\tau_1$) replicates the amplitude trend, decreasing gradually in the EI region, with small change in the semiconductor region, and subsequently experiencing another gradual decline in the semimetal region (Figure \ref{Fig:4p3}\textcolor{red}{b}). The slow relaxation amplitude A$_2$ behavior can be approximated with a systematic gradual fall upto P=1 GPa (Figure \ref{Fig:4p3}\textcolor{red}{c}). The corresponding relaxation time ($\tau_2$) depicts a linear fall in EI region (until P=1 GPa) followed by a constant value (within errorbars) for semiconductor as well as semimetal region for both the fluences (Figure \ref{Fig:4p3}\textcolor{red}{d}). 
\par
The relaxation dynamics of the condensed excitonic phase can be analyzed \cite{Werdehausen2018} by invoking the Rothwarf-Taylor (RT) model proposed in the context of superconductors\cite{Rothwarf1967}. The optical photoexcitation of excitonic insulator leads to the breaking of the excitons into unbounded electrons and holes. Following the photoexcitation, the excited electrons and holes recombine to form an exciton, accompanied by the release of high-frequency phonons (HFP). The recombination described here corresponds to the fast relaxation process (A$_1$, $\tau_1$). The HFP released in this process can either create another pair of electron and hole by breaking another exciton or decay into low frequency phonons (LFP), which will not be able to break any more excitons further. The former process can trigger an avalanche of free fermions as the transition point is approached, creating a bottleneck for the fermions to recombine into an exciton. This phenomenon results in a significant increase in relaxation time near the transition point\cite{Kabanov1999,Kabanov2005}. The latter process incorporates the fact that the decay of a HFP to low frequency phonons is much faster than the creation of fermion pair\cite{Torchinsky2010}. The fall of fast relaxation time ($\tau_1$), as the transition pressure (P=1 GPa) is approached (Figure \ref{Fig:4p3}\textcolor{red}{b}), elucidates the process where the HFP decays into LFPs. In addition to qualitative understanding, obtaining a quantitative estimate of the amplitude and relaxation time with pressure would offer more comprehensive insights.
\par
\begin{figure}[H]
    \centering
    \includegraphics[width=3.3in]{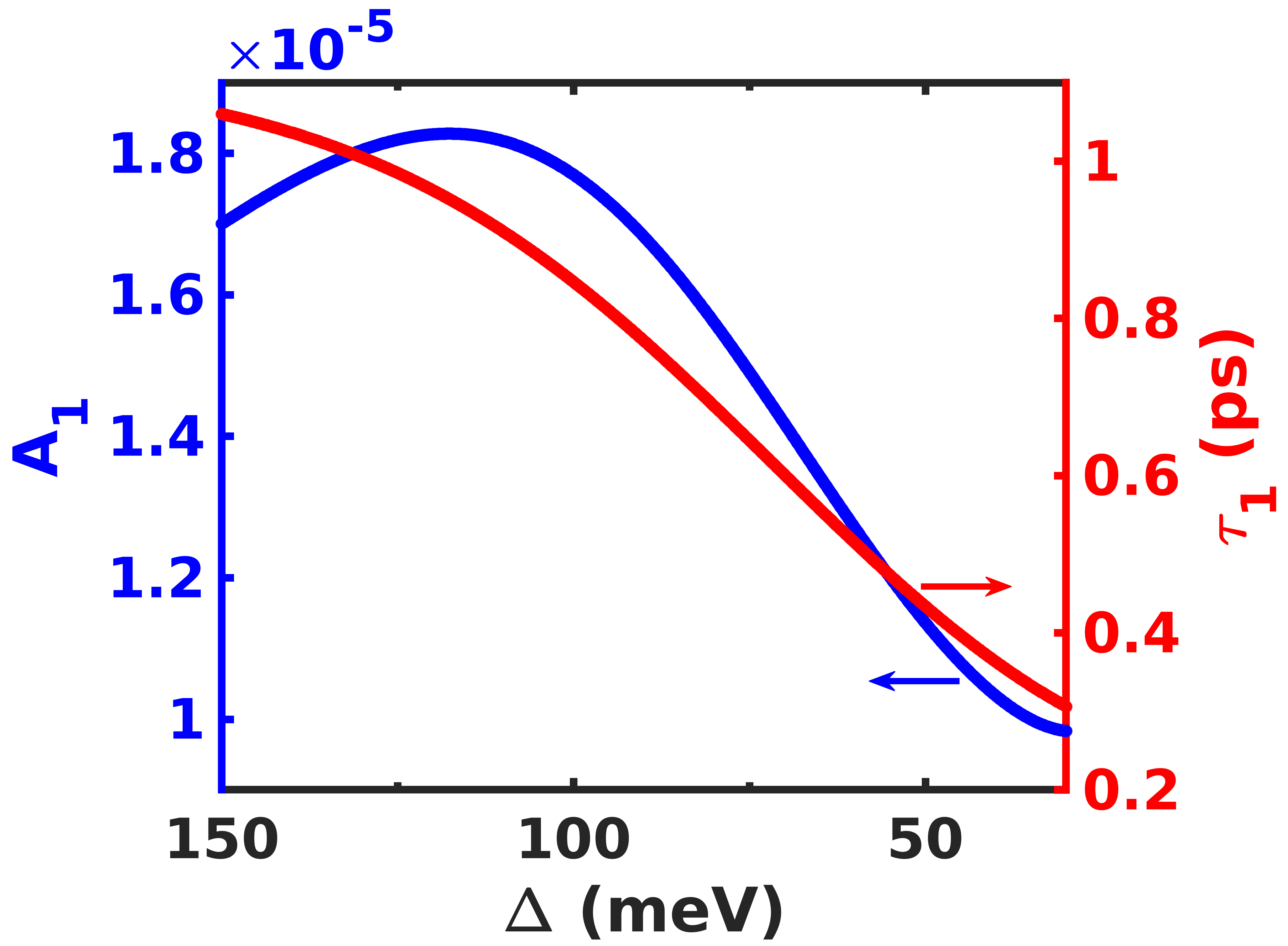}
    \caption[Amplitude and relaxation time with bandgap using Rothwarf-Taylor model]{Rothwarf Taylor model illustrates amplitude (A$_1$) and the relaxation time ($\tau_1$) (Eq.\ref{Eq2}) as a function of the bandgap.}
    \label{Fig:4p4}
\end{figure}

Exploiting the RT model, one can derive expressions for the amplitude (A$_1$) and the corresponding relaxation time ($\tau_1$), considering the temperature-independent gap (excitonic binding energy, $\Delta_{EI}\equiv2\Delta$)\cite{Kabanov1999,Kabanov2005,Werdehausen2018}.
\begin{align}
    A_1(T,P)=\frac{A_0/\Delta(P)}{1+Jexp(-\frac{\Delta(P)}{k_BT})} &&   \tau_1(T,P)=\frac{1}{K+L\sqrt{\Delta(P)T}exp(-\frac{\Delta(P)}{k_BT})}
    \label{Eq2}
\end{align}
where A$_0$, J, K,L and $\Delta$ are fitting parameters. 
In our experiments done at constant temperature (300 K), the variation of A$_1$ and $\tau_1$ arises due to pressure dependence of $\Delta$. The amplitude (A$_1$, blue curve) and relaxation time ($\tau_1$, red curve) are presented in Figure \ref{Fig:4p4} as a function of semi-bandgap ($\Delta$). Eq. \ref{Eq2} shows that A$_1$ and $\tau_1$ decrease with $\Delta$ (see Figure \ref{Fig:4p4}). At T=300 K and ambient pressure, the bandgap in the excitonic insulator is the binding energy of the excitons\cite{Lu2017}. In previous studies, it has been estimated to lie between 250 to 300 meV\cite{Tang2020,Wakisaka2012}. However, our calculations indicate a value, $\Delta_{EI}$ of 260 meV at room temperature and ambient pressure, which is a good approximation. At a pressure of P=1 GPa, EI phase becomes unstable, becoming a semiconductor with a bandgap 2$\Delta$ of 160 meV only\cite{Tang2020}. Using the Eq. \ref{Eq2} and the experimental value of A$_1$ and $\tau_1$ at different pressures, we evaluated the value of $\Delta$, the semi-bandgap. A linear reduction in $\Delta$ is observed with a slope of 65 meV/GPa, as the pressure is tuned from ambient to P=1 GPa. The pressure dependence of semi-bandgap, is expressed as $\Delta(P)$ (meV)$=\Delta_0-a*P$, where $\Delta_0=128$ meV and $a$=65 meV/GPa. This expression is utilized to evaluate the expression of amplitude (A$_1$) and the relaxation time ($\tau_1$) employing Eq. \ref{Eq2}, represented by the black solid curves in the EI phase of Figure \ref{Fig:4p3}\textcolor{red}{a} and \ref{Fig:4p3}\textcolor{red}{b}, respectively. These curves describe the trend of A$_1$ and $\tau_1$ with pressure (in the EI phase).
\par
To obtain pressure variation of the band gap in the EI phase from first principles, we simulated $2\times2 \times1$ supercell of optimized $C2/c$ structure of Ta$_2$NiSe$_5$ with $\Gamma$-point sampling. Next, we manually fixed the occupation of electron in KS-states at $\Gamma$ point such that an exciton is formed and estimated the total energy of the excited configuration of TNSe. Finally, the difference between the total energies of the excited and the ground states gives us the electronic band gap. This process is repeated for different pressures to get the pressure dependence of the electronic band gap of TNSe. The bandgap (2$\Delta$) obtained was 280, 200, and 140 meV under pressures of 0, 0.5, and 1.0 GPa respectively. The slope for the semi-bandgap variation, $\Delta$, is therefore 70 meV/GPa, closely aligning with the experimental value of 65 meV/GPa.

Comparing the amplitudes A$_1$ and A$_2$ (Figure \ref{Fig:4p3}) in the EI region, A$_1$ is five times than A$_2$, indicating the carrier dynamics being dominated by excitonic recombination. Similar to previous studies, we attribute the slow component in Eq. \ref{eq:4p1}, namely (A$_2$ and $\tau_2$), to thermalization of phonons (HFPs), emitted during the recombination of excitons\cite{Werdehausen2018}. The relaxation time $\tau_2$ represents the relaxation of hot phonons, which gradually cool down and decay into low-frequency phonons\cite{Werdehausen2018}. To understand the pressure dependence of $\tau_2$, we consider two approaches. In the first approach, we observe that the amplitude A$_2$ decreases with pressure. As the excitonic bandgap closes, there are fewer carriers available to recombine through the slow channel, resulting in a shorter recombination time. In another approach, the decrease in $\tau_2$ with pressure in the EI phase is similar to the increase in linewidth under pressure of the Raman modes in TNSe \cite{SukanyaPal2020}. Since the linewidth of a phonon is inversely related to its dephasing time \cite{Waldermann_2008}, this suggests a direct correlation between the recombination time and dephasing time, as reported in hybrid perovskites \cite{Zhang_2022}.
\par
In the semiconductor region spanning from P=1 GPa to 3 GPa, as illustrated in Figure \ref{Fig:4p3}, the relaxation dynamics is elucidated by the process of photoexcited carriers transferring energy to optical phonons, which subsequently relax through low-energy acoustic phonons. The fast relaxation time ($\tau_1$) corresponds to the relaxation of carriers by emitting optical phonons with a time scale of less than 1 ps, while the slow relaxation time ($\tau_2$) corresponds to the relaxation of acoustic phonons with a time scale of approximately 10 ps. We observe a pressure-independent behavior of both the relaxation times $\tau_1$ and $\tau_2$, as shown in Figure \ref{Fig:4p3}\textcolor{red}{b} and \ref{Fig:4p3}\textcolor{red}{d}, as expected because the phonon population is not influenced by pressure. Beyond 3 GPa, the fast relaxation time ($\tau_1$) decreases with pressure (Figure \ref{Fig:4p3}\textcolor{red}{b}). This is a consequence of the increasing carrier density in semi-metallic TNSe, leading to the availability of more relaxation channels. Understanding the phenomenon under different excitation densities is helpful for describing the various phases of TNSe, especially considering the role of electron-phonon coupling in the process.
\par
\begin{figure}[H]
    \centering
    \includegraphics[width=\textwidth]{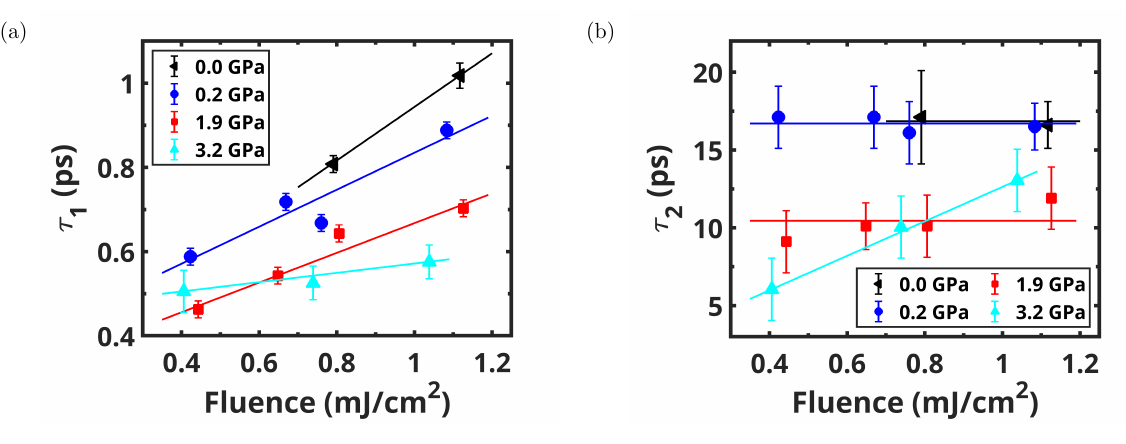}
    \caption[Fluence dependence of relaxation times at different pressures]{(a) $\tau_1$ versus pump fluence at different pressures (b) $\tau_2$ vs fluence at different pressures. The solid lines are linear fits.}
    \label{Fig:4p5}
\end{figure}

Figure \ref{Fig:4p5}\textcolor{red}{a} illustrates the variation of $\tau_1$ with fluence at different pressures, revealing a direct relationship between relaxation time and excitation density. With a fixed phonon population density at a given pressure and temperature, an increase in fluence results in a higher density of photoexcited carriers and hence, leading to an increase in relaxation time. The slope of the increment in $\tau_1$ with fluence is minimum at the pressure of 3.2 GPa, highlighting the semimetal region characterized by a higher carrier density and the presence of additional relaxation channels. The slow relaxation time, $\tau_2$, is observed to be independent of fluence below 3 GPa, as depicted in Figure \ref{Fig:4p5}\textcolor{red}{b}. This indicates that sufficient low-frequency acoustic phonons are available for the maximum excitation density ($\sim$1.1 mJ/cm$^2$) to facilitate the decay of high-frequency optical phonons. However, the increase of $\tau_2$ with fluence above 3 GPa can be attributed to the phonon bottleneck effect.

\section*{Conclusions}
Through the application of on-site, in-situ high-pressure optical pump-probe spectroscopy on TNSe, we were able to elucidate the distinct phases of the material. TNSe exhibits excitonic insulator behavior in the pressure range of 0 to 1 GPa, transitions to a semiconductor phase between 1 to 3 GPa, and transforms into a semimetallic state beyond 3 GPa. By incorporating the Rothwarf-Taylor model for the EI phase, we have demonstrated that pressure linearly tunes the bandgap as the transition pressure is approached. The slope of the semibandgap variation with pressure was evaluated to be 65 meV/GPa, consistent with the value of 70 meV/GPa calculated by first principles. The slow relaxation time, associated with the decay of hot phonons, decreases with increasing pressure. A direction correlation is observed between carrier recombination time and phonon dephasing time of phonon as a function of pressure in the EI phase. The semiconductor and semimetallic phases of TNSe can be explained by the electron-phonon interaction. This study not only illustrates the pressure tuning of the bandgap in excitonic insulators but also provides insights into the microscopic electron-phonon interactions under high-pressure conditions.

\section*{Acknowledgments} 
AKS thanks the Department of Science and Technology under the National Science Chair Professorship for financial support, and VA acknowledges CSIR for the research fellowship.


\floatsetup[figure]{style=plain,subcapbesideposition=top}


\clearpage
\newpage
\bibliographystyle{unsrt}
\bibliography{citations_Ta2NiSe5_P.bib}

\end{document}